\documentclass[%
reprint,
superscriptaddress,
 amsmath,amssymb,
 aps,
]{revtex4-2}

\usepackage{graphicx}
\graphicspath{ {Figures/} }
\usepackage{dcolumn}
\usepackage{bm}
\usepackage[hyperindex,breaklinks]{hyperref}
\hypersetup{colorlinks=true}
\usepackage{breakurl}
\usepackage[mathlines]{lineno}
\usepackage[caption=false]{subfig}

\newcommand{\nuebar}{$\overline{\nu}_{e}$ }

\newcommand{\alphan}{($\alpha, n$) }

\begin{document}

\preprint{APS/123-QED}

\title{Initial measurement of reactor antineutrino oscillation at SNO+}

\author{ A.\,Allega}
\affiliation{\it Queen's University, Department of Physics, Engineering Physics \& Astronomy, Kingston, ON K7L 3N6, Canada}
\author{ M.\,R.\,Anderson}
\affiliation{\it Queen's University, Department of Physics, Engineering Physics \& Astronomy, Kingston, ON K7L 3N6, Canada}
\author{ S.\,Andringa}
\affiliation{\it LIP - Laborat\'{o}rio de Instrumenta\c{c}\~{a}o e  F\'{\i}sica Experimental de Part\'{\i}culas, Av. Prof. Gama Pinto, 2, 1649-003 Lisboa, Portugal}
\author{ M.\,Askins}
\affiliation{\it University of California, Berkeley, Department of Physics, CA 94720, Berkeley, USA}
\affiliation{\it Lawrence Berkeley National Laboratory, 1 Cyclotron Road, Berkeley, CA 94720-8153, USA}
\author{ D.\,M.\,Asner}
\affiliation{\it Brookhaven National Laboratory, P.O. Box 5000, Upton, NY 11973-500, USA}
\author{ D.\,J.\,Auty}
\affiliation{\it University of Alberta, Department of Physics, 4-181 CCIS,  Edmonton, AB T6G 2E1, Canada}

\author{ A.\,Bacon}
\affiliation{\it University of Pennsylvania, Department of Physics \& Astronomy, 209 South 33rd Street, Philadelphia, PA 19104-6396, USA}
\author{J.\,Baker}
\affiliation{\it SNOLAB, Creighton Mine \#9, 1039 Regional Road 24, Sudbury, ON P3Y 1N2, Canada}
\author{ F.\,Bar\~{a}o}
\affiliation{\it LIP - Laborat\'{o}rio de Instrumenta\c{c}\~{a}o e  F\'{\i}sica Experimental de Part\'{\i}culas, Av. Prof. Gama Pinto, 2, 1649-003 Lisboa, Portugal}
\affiliation{\it Universidade de Lisboa, Instituto Superior T\'{e}cnico (IST), Departamento de F\'{\i}sica, Av. Rovisco Pais, 1049-001 Lisboa, Portugal}
\author{ N.\,Barros}
\affiliation{\it LIP - Laborat\'{o}rio de Instrumenta\c{c}\~{a}o e  F\'{\i}sica Experimental de Part\'{\i}culas, Rua Larga, 3004-516 Coimbra, Portugal}
\affiliation{\it Universidade de Coimbra, 
Departamento de F\'{\i}sica (FCTUC), 3004-516, Coimbra, Portugal}
% ...

\collaboration{The SNO+ Collaboration}


\begin{abstract}
\newpage

The SNO+ collaboration reports its first spectral analysis of long-baseline reactor antineutrino oscillation using 114 tonne-years of data. 
Fitting the neutrino oscillation probability to the observed energy spectrum yields constraints on the neutrino mass-squared difference $\Delta m^2_{21}$. 
In the ranges allowed by previous measurements, the best-fit $\Delta m^2_{21}$ is (8.85$^{+1.10}_{-1.33}$) $\times$ 10$^{-5}$ eV$^2$.  
This measurement is continuing in the next phases of SNO+ and is expected to surpass the present global precision on $\Delta m^2_{21}$ with about three years of data.  

\end{abstract}
 
\maketitle


\section{\label{sec:intro}Introduction}

Reactor antineutrino experiments have produced leading measurements of neutrino oscillation parameters $\Delta m^2_{21}$, $\theta_{13}$, and $\Delta m^2_{32}$~\cite{Eguchi:2002dm, Dayabay:2012dis} and are expected to produce more precise measurements in the near future~\cite{An:2015jdp}. The current precision on $\Delta m^2_{21}$ is dominated by the lone measurement of long-baseline reactor antineutrinos from the KamLAND experiment~\cite{KamLAND:2013rgu}. 
The analysis of all solar neutrino experiments by Super-K results in a value that is in 1.5$\sigma$ tension~\cite{Super-Kamiokande:2023jbt}.  
Thus, additional precise measurements using reactor or solar neutrinos are of interest. 

Reactor antineutrinos are detected via inverse beta decay (IBD) on hydrogen: 
$\overline{\nu}_e + p \rightarrow e^+ + n$, which has a 1.81-MeV threshold.  The $e^+$ carries most of the energy from the \nuebar and the $n$ subsequently thermalizes, finally producing a 2.22-MeV $\gamma$ when it captures on a hydrogen nucleus.  
The SNO+ collaboration recently reported the first evidence of reactor \nuebar in a large water Cherenkov detector~\cite{SNO:2023wan}, also identifying IBDs with neutron captures on hydrogen.  In that measurement, the 2.22-MeV $\gamma$ was only partially above the detector energy threshold and random coincidences of ambient radioactivity were a major background. 
Furthermore, the relatively poor energy resolution of Cherenkov detectors at MeV energies diminishes the ability to observe spectral features from neutrino oscillation.  

The SNO+ collaboration reports here its first measurement of reactor antineutrino oscillation, using liquid scintillator.  
The higher light yield of the scintillator provides finer energy resolution, which enables the study of spectral features due to neutrino oscillation, as well as a better discrimination of signals from backgrounds.  
Additionally, the levels of radioactivity in the SNO+ scintillator are one to two orders of magnitude lower than in the SNO+ water, further decreasing the random coincidences.  
As a result, the dominant background for the current analysis is from $^{13}$C($\alpha, n$)$^{16}$O reactions in the scintillator, which was also the case for the measurements from KamLAND~\cite{KamLAND:2013rgu}.  

In the following, we first describe the configuration of the SNO+ detector when it was partially filled with scintillator, and a characterization of the detector response using intrinsic radioactivity.  Next, we detail the event selection and expectations for reactor IBDs and \alphan reactions.  Then, we present the results of an energy spectrum analysis using 114 tonne-years of data.  
We conclude with prospects of future results 
from the SNO+ detector, which has been operating fully-filled with 780 tonnes of  scintillator.

\section{Data}
\begin{figure}[htbp]
\caption{\label{fig:detector}
Picture of the detector during scintillator fill from an underwater camera mounted next to the PMTs. The horizontal scintillator-water interface is clearly visible, well below the AV neck. 
}
\includegraphics[width=\columnwidth]{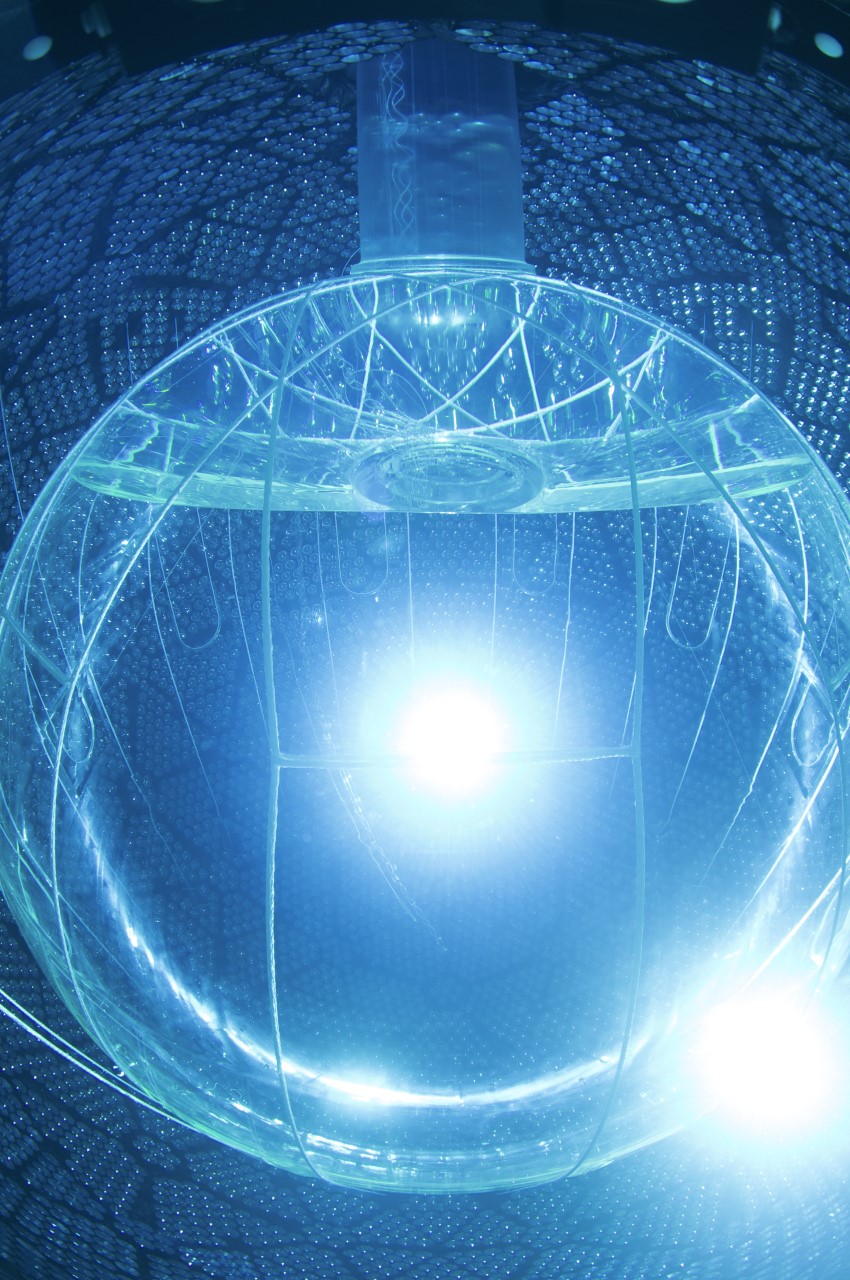}
\end{figure}

SNO+ is a multipurpose neutrino experiment located 2~km underground in Ontario, Canada.  The detector consists of an acrylic vessel (AV) with a 6.0-m radius that is surrounded by ultrapure water and 9362 photomultiplier tubes (PMTs) at a radius of about 8.5~m.  At the top of the AV, a cylindrical `neck' of 1.5-m diameter extends 7~m upward.  The detector is depicted in Fig.~\ref{fig:detector} and described in detail in Ref.~\cite{SNO:2021xpa}. 

From September 2017 to July 2019, the AV was filled with 905 tonnes of ultrapure water and the SNO+ detector operated as a low-threshold water Cherenkov detector. 
After this water phase, the collaboration filled the detector with liquid scintillator, injecting near the top of the AV neck while extracting water from the bottom of the AV.  
Because of the COVID-19 pandemic, scintillator filling was paused between March and October 2020, with the scintillator at a height of about 75~cm above the AV equator, providing 130.2 days of stable data that is used in this work.  In this partial-fill phase, the spherical volume of the AV contained about 320 tonnes 
of linear alkyl benzene (LAB), plus the fluor 2,5-diphenyhloxazole (PPO) at a concentration of 0.6~g/L.  
Figure~\ref{fig:detector} shows a photo of the detector in which the horizontal interface between the scintillator and water is clearly seen during an earlier stage of filling.

The vertical offset between the geometric centers of the PMT support structure and the AV was measured to be 13~cm, and is taken into account in simulations. Events originating from the scintillator in the neck region beyond the PMTs are rejected using simple geometrical algorithms.  
The detector optics were extensively calibrated and modeled during the water phase~\cite{SNO:2021wcl}. For the partial-fill phase, the scintillator characteristics are modeled based on ex-situ measurements~\cite{SNO:2020fhu} and the model is empirically tuned using naturally occurring radioactivity, as explained in the next section. 

We refer to a PMT that detects one or more photons as a hit PMT.  
The detector trigger threshold was set to approximately 10 hit PMTs, corresponding to roughly 40~keV for an electron. 
For analysis, valid hit PMTs are selected using the same criteria as in the water phase~\cite{pmtsel} and the event position is reconstructed using a similar likelihood fit based on time-of-flight-corrected PMT hit times. 
The energy of an event is reconstructed largely based on the number of hit PMTs, but also accounts for photon propagation and PMT detection efficiency, given the reconstructed event position.  Events above a few MeV are likely to result in some PMTs detecting multiple photons. 
This effect is negligible for 1-MeV electrons, but for 10-MeV electrons, around 25\% of detected photons are incident on already-hit PMTs.  
The energy reconstruction accounts for this based on the observed spatial distribution of hit PMTs. 
About 20\% of the dataset was acquired with one electronics crate off, which excludes 512 PMT channels along a vertical wedge of the spherical detector. This results in fewer hit PMTs for the same deposited energy, which is accounted for in simulation and, to first order, in the energy reconstruction. 

Most detected events were due to ambient radioactivity from the uranium and thorium decay chains. 
From studies of Bi-Po sequential decays, the equilibrium concentrations of $^{238}$U and $^{232}$Th in the scintillator are both estimated to be 5 $\times$ 10$^{-17}$~g/g$_{\mathrm{scint}}$.  
These concentrations can be compared to the initial values in KamLAND, which were 0.35 $\times$ 10$^{-17}$ and 5.2 $\times$ 10$^{-17}$~g/g$_{\mathrm{scint}}$, respectively~\cite{Eguchi:2002dm}.  
Of main interest for this work are the 5.3-MeV $\alpha$'s from $^{210}$Po decay, which can induce \alphan reactions.

\section{Scintillator calibration}

The scintillator characteristics during the partial-fill phase were based on ex-situ measurements~\cite{{SNO:2020fhu}} and further tuned using the sequential decays of $^{214}$Bi ($\beta$; 3.3-MeV Q-value) and $^{214}$Po ($\alpha$; 7.8-MeV Q-value and 164~$\mu s$ half-life).  
This coincidence background was present in large quantities while filling the detector, due to $^{222}$Rn entering the detector through the AV neck and liquid circulation systems, and reduced at later stages of data-taking as the $^{222}$Rn decayed with a halflife of 3.8~days.
The full dataset was analyzed within the fiducial volume defined as 85~cm above the equator (10~cm above the water level) and within a 5.7-m radius (30~cm from the AV). With the simple set of coincidence criteria summarized in Table~\ref{tab:cuts}, pure samples of $\beta$'s and $\alpha$'s are selected with an estimated contamination of the order of $10^{-3}\%$.  The scintillation signal of $\alpha$'s is quenched by an order of magnitude relative to $\beta$'s, resulting in fewer hit PMTs for the same kinetic energy.  

\begin{table}[thbp]\centering
\caption{Basic selection criteria for coincidence events. 
 $R$ and $Z$ represent radial and vertical coordinates.  Time and position differences are between prompt and delayed events.  
 }\label{tab:cuts}
\begin{tabular}{l|cc|cc}
\hline
\hline
& \multicolumn{2}{c|}{Calibration} & \multicolumn{2}{c}{Antineutrino} \\ 
  & $^{214}$Bi $\beta$ & $^{214}$Po $\alpha$ & IBD $e^+$ & IBD $n$ \\
\hline
Fid. volume [m]& \multicolumn{2}{c|}{($Z>0.85$, $R<5.7$)} & \multicolumn{2}{c}{($Z>0.85$, $R<5.7$)} \\
 Time diff. [$\mu$s] & \multicolumn{2}{c|}{[0.4, 1000]} & \multicolumn{2}{c}{[0.4, 800]} \\ 
Position diff. [m] & \multicolumn{2}{c|}{$<1.0$} & \multicolumn{2}{c}{$<1.5$} \\ 
Hit PMTs & [330,1050] & [170,320] & -- & -- \\
Energy [MeV]& -- & -- & [0.9,8.0] & [1.85,2.40] \\
\hline
\hline
\end{tabular}
\end{table}

\subsection{Scintillation time profile}
The scintillation time profile is different for $\alpha$'s and $\beta$'s, and is reflected in the distribution of the time-of-flight-corrected PMT hit times. 
In the simulations, these profiles are parameterized by a shared exponential rise component (0.8~ns) and three particle-dependent exponential decay components. The exponential decay parameters are scanned to find the best match with the BiPo calibration data~\cite{SNO:2023cnz}.  As expected, $\alpha$'s exhibit a slower emission time profile.  This calibration also results in a better agreement between data and simulation in the inter-event distance between the $\alpha$ and $\beta$ signals~\cite{IMB2021}. 

\subsection{Scintillation yield and quenching}
The $^{214}$Bi $\beta$ decay energy spectrum is used to measure the energy scale, 
which is found to be 330 hit PMTs/MeV at the detector center. The scintillator light yield in simulation is set to 6694 photons/MeV in order to match the number of hit PMTs observed in data. 
After this calibration, the $^{214}$Po $\alpha$ decay energy peak still showed a difference between data and simulation. This is empirically corrected in simulation by setting the Birks' constant for $\alpha$'s to 80.3~$\mu$m/MeV, while keeping that for $\beta$'s at 79.8~$\mu$m/MeV.~\cite{IMB2021}

\subsection{Energy scale uniformity}
After light yield and quenching adjustments, a residual position dependence of the energy scale is still present near the AV. 
This was corrected empirically by matching the medians of the energy distributions in the BiPo data and simulations at various regions of vertical position $Z$ and horizontal radius $\rho \equiv \sqrt{X^2 + Y^2}$.  The correction is consistent between $\alpha$'s and $\beta$'s~\cite{IMB2021}.  After applying the correction, the energy scale and resolution are compatible between data and simulation, within a statistical uncertainty of 3\% across both $Z$ and $\rho$, for both $\alpha$'s and $\beta$'s.  This is taken as a systematic uncertainty, and is much smaller than the statistical uncertainty in the present analysis. 
The energy resolution at the detector center is about 6\% at 1~MeV.

\section{Signals and Backgrounds}\label{sec:sigBg}

Reactor antineutrinos are selected as time coincidences of a prompt event with energy between 0.9~MeV and 8.0~MeV, and a delayed event in the range of 1.85~MeV to 2.40~MeV.  These two ranges select the positron and 2.22-MeV neutron-capture $\gamma$'s with high efficiency.  
The coincidence time window is [0.4, 800]~$\mu$s to ensure a high efficiency to identify the neutrons, which have a mean capture time around 210~$\mu$s in both the scintillator and water.
These selection criteria are shown in Table~\ref{tab:cuts} and are determined by simulation to have an efficiency of 78\% for reactor IBDs in the fiducial volume, which comprises 90\% of the scintillator.

To avoid muon spallation products, such as neutrons and $\beta$-$n$ decaying isotopes like $^9$Li, the 20~s of data after any event with more than 3000 hit PMTs ($\approx$10~MeV) are excluded from analysis.  
Similarly, to avoid contamination from neutron-producing atmospheric neutrinos, 
data are vetoed within $\pm$2~ms around a prompt event candidate when more than one delayed event candidate is observed within 2~m of the prompt position.  No coincidences in the current dataset are rejected by these criteria while the livetime is reduced by 3.7\%, to 125.4 days.  

\begin{figure}[thbp]
\caption{\label{fig:IBDselection}
Time between prompt and delayed events (top); Distance between prompt and delayed events (bottom). The exponential fit results in $253\pm52~\mu$s, consistent with the expectation for neutron capture. The IBD simulations are normalized to the 45 observed events.  Error bars are Poisson.}
\includegraphics[width=\columnwidth]{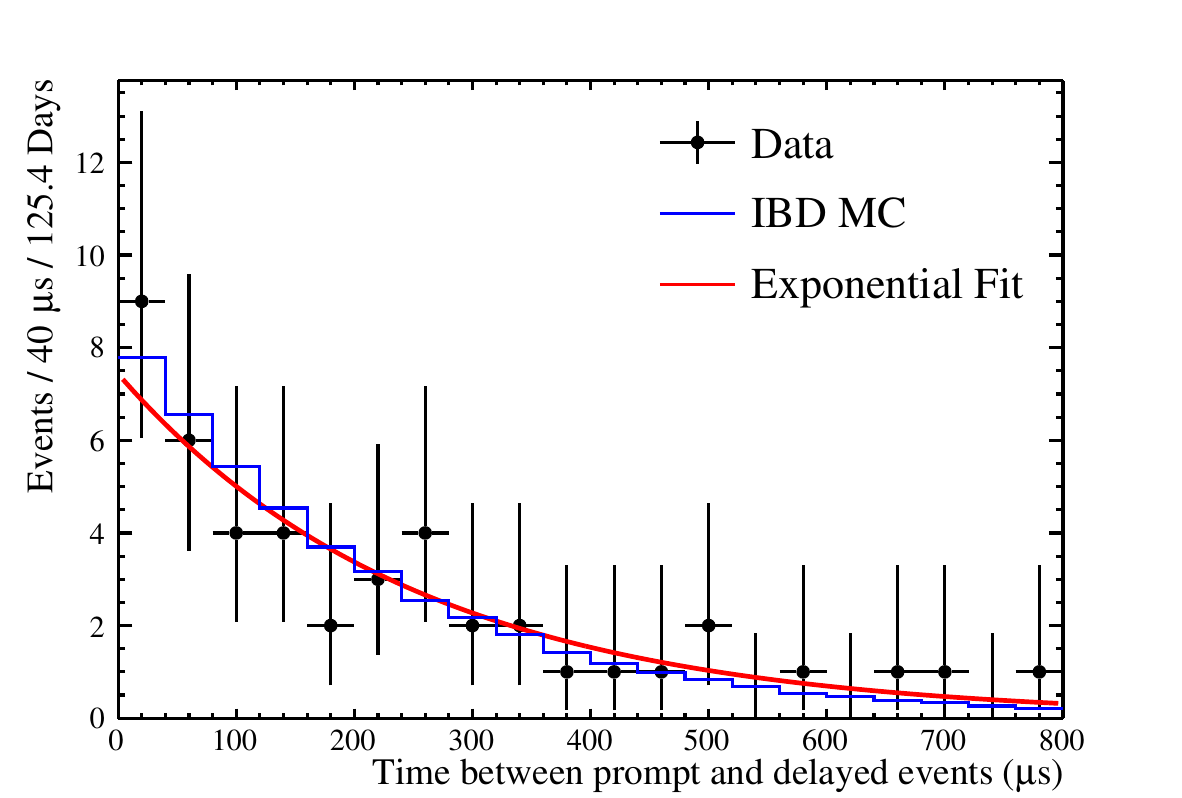}
\includegraphics[width=\columnwidth]{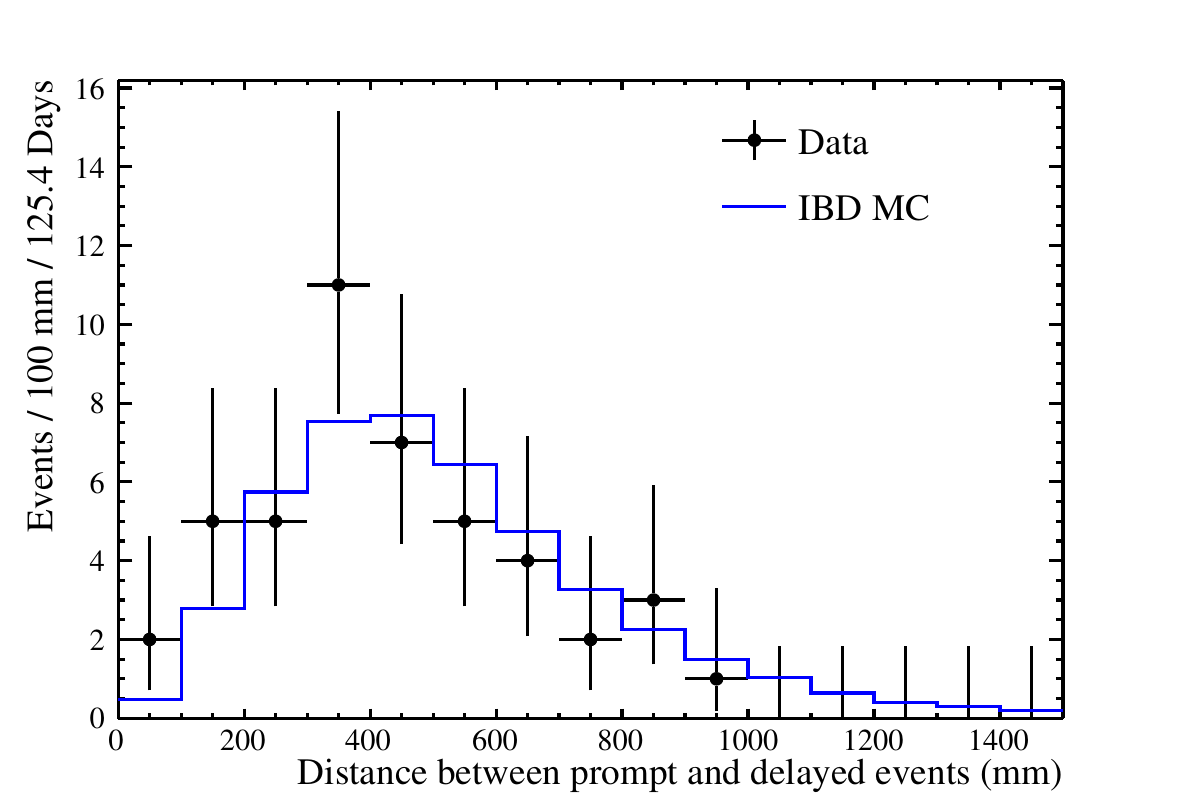}
\end{figure}

After all selection criteria are applied, 45 coincidences are observed in the data, and are listed in Table~\ref{tab:event_info}.  Figure~\ref{fig:IBDselection} compares the data with the associated IBD simulations, showing that the sample is a pure selection of coincidences with delayed neutron captures.  The individual expectations for the antineutrino signals and backgrounds are discussed below. More details can be found in Ref.~\cite{IMB2021}.

\subsection{Reactor antineutrinos} 

The predicted rate and energy spectrum of reactor IBDs uses the Huber-Mueller isotope model and other inputs, as described in Ref.~\cite{DayaBay:2016ssb}.  
Nearly 60\% of the IBDs in the SNO+ detector originate from three Canada Deuterium Uranium (CANDU) reactor complexes, at distances of 240, 340, and 350~km.  
The remaining flux originates from approximately 100 cores in the USA.  

Electron antineutrinos are produced at reactors at a rate of 2 $\times$ 10$^{20}$ per second per GW of thermal power.  Thermal powers for each reactor core are obtained from monthly averages provided by the IAEA~\cite{PRIS}.  The three CANDU reactor complexes are modeled using hourly electrical power provided by IESO~\cite{IESO}.  Averaging these values and comparing with the IAEA results reveals a difference of (+0.2$\pm$0.1)\% across a 12-month period.  

The \nuebar flux and energy spectrum also depend on the relative fractions of fissile isotopes, which evolve with time.  The incident \nuebar flux varies by less than 1\% because of the large number of cores, of which CANDU reactors are constantly refueled.  Therefore, average fission fractions are used in the predictions for the CANDU pressurized heavy water reactors (PHWRs), as well as pressurized/boiling water reactors (P/BWRs):
($^{235}$U, $^{239}$Pu, $^{238}$U, $^{241}$Pu) are set to
(0.52, 0.42, 0.05, 0.01) for PHWRs~\cite{AECL:2013} and to 
(0.568, 0.297, 0.078, 0.057) for P/BWRs~\cite{Eguchi:2002dm}.  
The latter values agree to within 1\% (absolute) with the values from Ref.~\cite{PRD5}.  From this same reference, the uncertainty of the fission fractions propagates to a 0.6\% uncertainty on the flux. 

When compared with past reactor IBD measurements, the flux of the Huber-Mueller isotope model has been shown to be biased and is corrected by scaling it to the global average of reactor flux measurements, i.e., multiplying by 0.945$\pm$0.007~\cite{DayaBay:2018heb}. 
The spectrum of the isotope model has also been shown to be discrepant and predicted to introduce a roughly 2.7\% uncertainty on the flux from P/BWRs~\cite{DayaBay:2016ssb}.  
Systematic uncertainty components are largely taken from Ref.~\cite{DayaBay:2018heb} and total to around $\pm$3\% on the rate of IBDs, which is negligible relative to the statistical uncertainty of the current dataset.  

The survival probability of an electron (anti)neutrino is 
\begin{equation*}
\begin{aligned}
P_{ee} = 1
& \left. -\cos^{4}\theta_{13}\sin^{2}2\theta_{12}\sin^{2}\Delta_{21}\right.\\
& \left. -\sin^{2}2\theta_{13}(\cos^{2}\theta_{12}\sin^{2}\Delta_{31} + \sin^{2}\theta_{12}\sin^{2}\Delta_{32}) \right.\\
  \approx (1
& -\sin^{2}2\theta_{12}\sin^{2}\Delta_{21})\cos^{4}\theta_{13}+\sin^4\theta_{13},
\end{aligned}
\end{equation*}
where $\Delta_{ij} \equiv 1.267\Delta m_{ij}^{2}L/E$, $E$ [MeV] is the energy of the neutrino, $L$ [m] is the distance traveled by the neutrino, and $\Delta m_{ij}^{2} \equiv m_i^2-m_j^2$ [eV$^2]$ is the difference between the squares of the masses of neutrino mass eigenstates $i$ and $j$.  
The approximation, which is accurate at a few MeV and several hundreds or thousands of kilometers, helps illustrate how $\Delta m_{21}^{2}$ and $\theta_{12}$ are the dominant parameters in determining the energy spectrum and rate of reactor IBDs, respectively.  As a result, the choice of neutrino mass ordering has negligible impact.  
With input values from Ref.~\cite{PDG2021}, assuming the normal neutrino mass ordering, the fraction of IBDs at SNO+ will be reduced, to about $<P_{ee}>~=~0.55$. This value increases by less than 1\% when the matter effect of the Earth's crust is included in the calculation~\cite{Page:2023rpb}.  

\begin{figure}[tpb]
\caption{\label{fig:EpromptPredict}
Predicted energy spectra of reactor IBD prompt events assuming no oscillation (black), $\Delta m^2_{21}=7.53\times10^{-5}$~eV$^2$ from KamLAND~\cite{KamLAND:2013rgu} (blue), $\Delta m^2_{21}=6.10\times10^{-5}$~eV$^2$ from solar measurements~\cite{Super-Kamiokande:2023jbt} (red), and $\Delta m^2_{21}=8.85\times10^{-5}$~eV$^2$ from the present analysis (green).}
\includegraphics[width=\columnwidth]{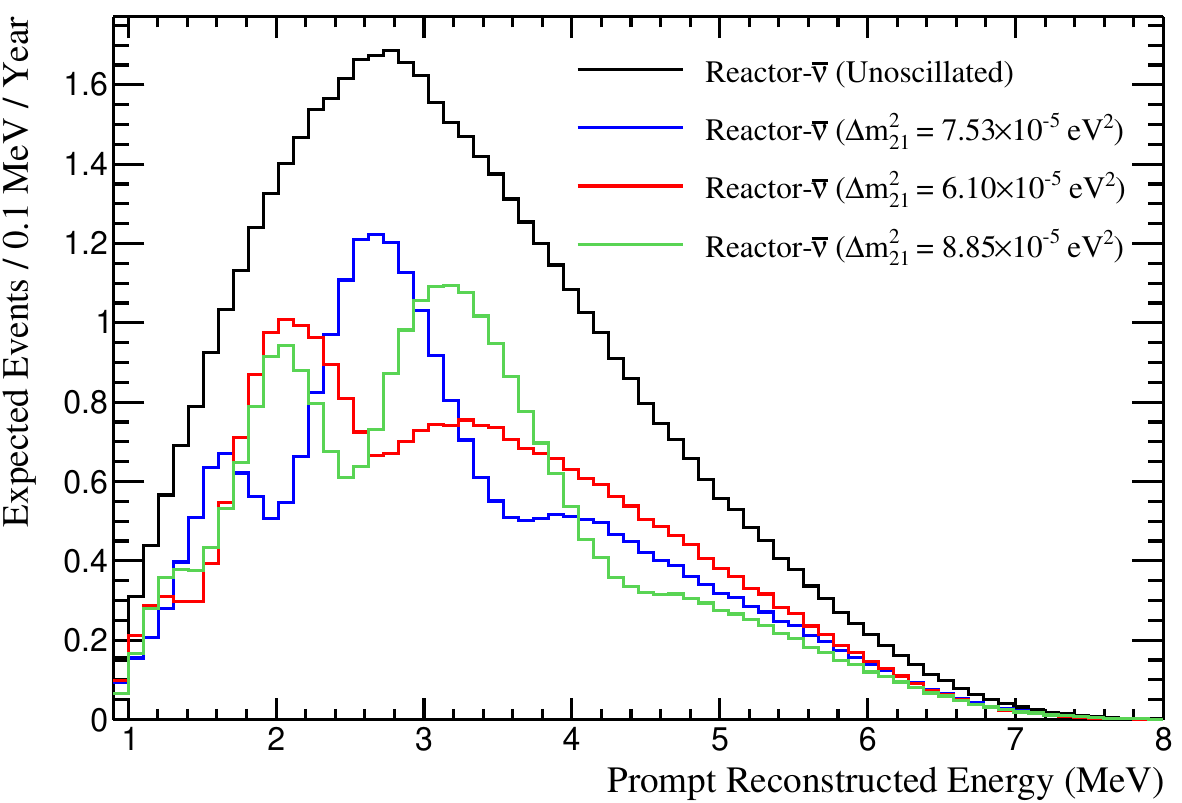}
\end{figure}

For $\sin^{2}\theta_{12}=0.307$ and $\Delta m^2_{21}=7.53\times10^{-5}$~eV$^2$~\cite{PDG2021}, around 100 IBDs are expected per year when the volume enclosed by the AV is completely filled with scintillator.  Of these, 40 IBDs would come from the nearest reactor complex at 240~km and 20 IBDs from about 350~km away, providing clear oscillation patterns in the measured energy spectrum, as seen in Fig.~\ref{fig:EpromptPredict}.  
Taking into account the 78\% selection efficiency and the fiducial volume, this translates to an expectation of 9.53$\pm$0.30 reactor IBDs for the present dataset of 125.4 days.  

\subsection{Geoneutrinos}
The uranium and thorium decay chains present in the Earth also produce antineutrinos above the IBD threshold of 1.81~MeV. 
The flux of these geoneutrinos strongly depends on the local geology and the geological model used.  
An estimate based on the method of Ref.~\cite{geo}, and assuming 20 TW of radiogenic heat, gives 34.1$\pm$5.0 and 9.5$\pm$0.8 Terrestrial Neutrino Units (TNU) from $^{238}$U and $^{232}$Th, respectively.  TNU is defined as one IBD interaction in one year of fully efficient exposure to $10^{32}$ free protons. This leads to an expectation of 2.2 selected IBDs in the present dataset.  
The predicted energy spectrum corresponding to these components is shown in Fig.~\ref{fig:Eprompt}. The impact of neutrino oscillations is included simply as a multiplicative constant of $<P_{ee}>~=~0.55$. 
Recognizing the wide range of possible heat values and variations in local geology, we assign a 100\% systematic uncertainty to the total prediction.

\subsection{\alphan backgrounds}

The prompt signals of \alphan interactions are from (1) protons scattered by the neutron (plus a small contribution from the energy deposited by the $\alpha$ itself), or (2) an excited state emitting a 6-MeV $\gamma$ or $e^+ e^-$ pair, or (3) the neutron exciting $^{12}$C, which then emits a 4.4-MeV $\gamma$. 
The predicted energy spectrum corresponding to these components is shown in Fig.~\ref{fig:Eprompt}.  

In the water phase, the rate of \alphan induced by the $\alpha$ decay of $^{210}$Po on the AV was measured from interactions with $^{13}$C and $^{18}$O that resulted in excited states of $^{16}$O and $^{21}$Ne~\cite{SNO:2023wan}. The background of \alphan from the AV or the AV-external water is reduced by the fiducial volume selection, but \alphan interactions with $^{13}$C now occur within the AV-internal volume, in the scintillator. 

The rate of $^{210}$Po $\alpha$ decays is now directly measured in the fiducial volume by fitting the $\alpha$ energy peak centered around 0.4~MeV (quenched down from 5.3~MeV).  The average rate over the this dataset is 85~Hz, which is several orders of magnitude greater than the rate of $^{214}$Po $\alpha$ decays. 
An expected \alphan rate of 5.27~$\mu$Hz is then obtained from the cross-section~\cite{JENDL150}, $\alpha$ energy loss in propagation~\cite{srim}, and the number density of $^{13}$C in the SNO+ scintillator.  
To reflect disagreements between the parameterized cross-section and direct measurements~\cite{Harissopulos153}, we assign a 30\% uncertainty to the dominant ground state signal, and a 100\% uncertainty to the two excited state signals, which together have a 9.2\% branching ratio.
\alphan are the major background for the present analysis, with a prediction of 33.3$\pm$12.7 selected coincidences.

\subsection{Other backgrounds}

Neutral current interactions of atmospheric neutrinos can also result in delayed neutron captures associated with prompt interactions. This was one of the backgrounds in the SNO+ water phase, due to the de-excitation signal of $^{15}$O*. Repeating the same simulation study with the partially-filled scintillator detector yielded a prediction of $< 1$ selected coincidence in the present dataset.  

Fast neutrons and ($\beta$+$\gamma$, $n$) reactions induced by cosmogenic muons are also considered, but are small in relation to the $^{210}$Po-induced \alphan contribution expected in the dataset.  
In addition, all muon products are found to be negligible after the exclusion of 20~s of data following muons. 

Random coincidences from ambient radioactivity are estimated by using the measured rates of prompt and delayed candidates, and randomly pairing events before applying the coincidence cuts. The calculation is checked by using a much larger time window, outside the IBD coincidence window.  The expected number of random coincidences selected is 0.216 $\pm$ 0.002, and their contribution is not considered in the following analysis.

\section{Results}

\begin{figure}[hbtp]
\caption{\label{fig:Eprompt} Energy distributions of prompt events.  
Oscillation parameters used in reactor IBD prediction are from Ref.~\cite{PDG2021}, assuming normal neutrino mass ordering.  Error bars are Poisson.  Data are given wider bins to aid visual comparison.}
\includegraphics[width=\columnwidth]{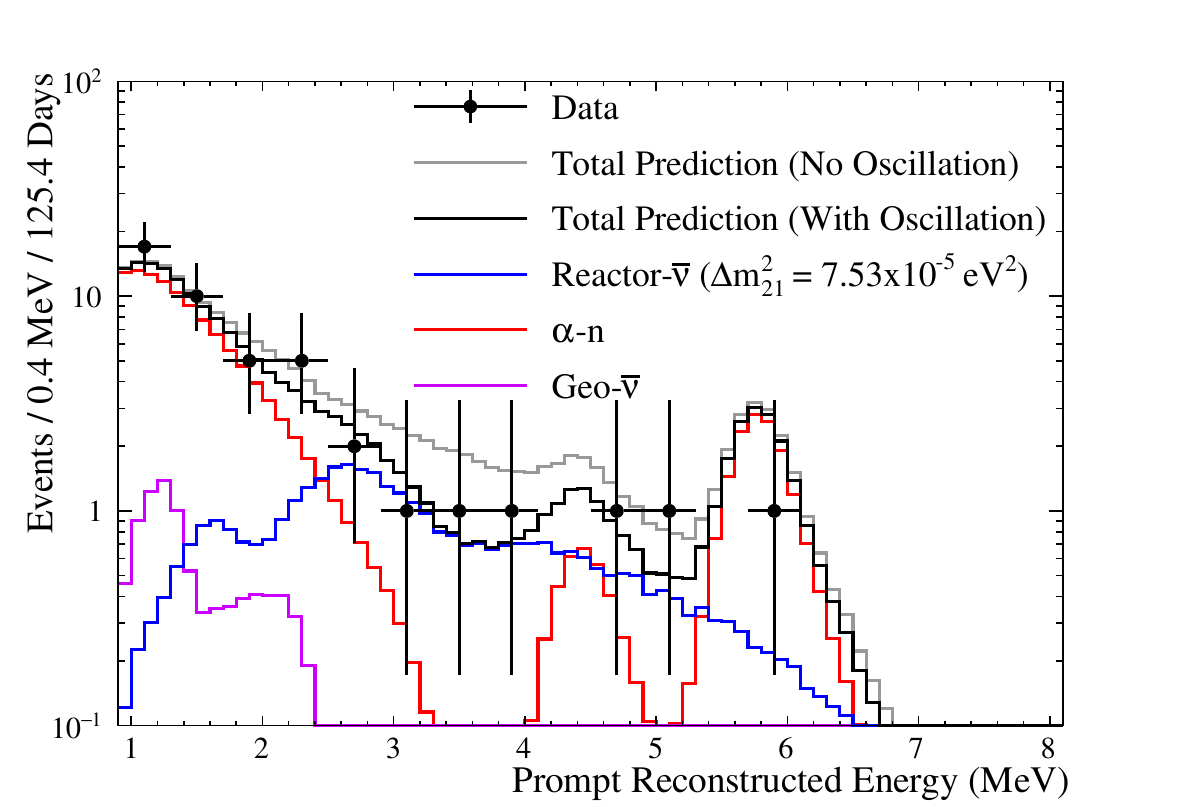}
\end{figure}

\begin{figure}[hbtp]
\caption{\label{fig:ZvsRho} Spatial distribution of observed coincidence pairs.}
\includegraphics[width=\columnwidth]{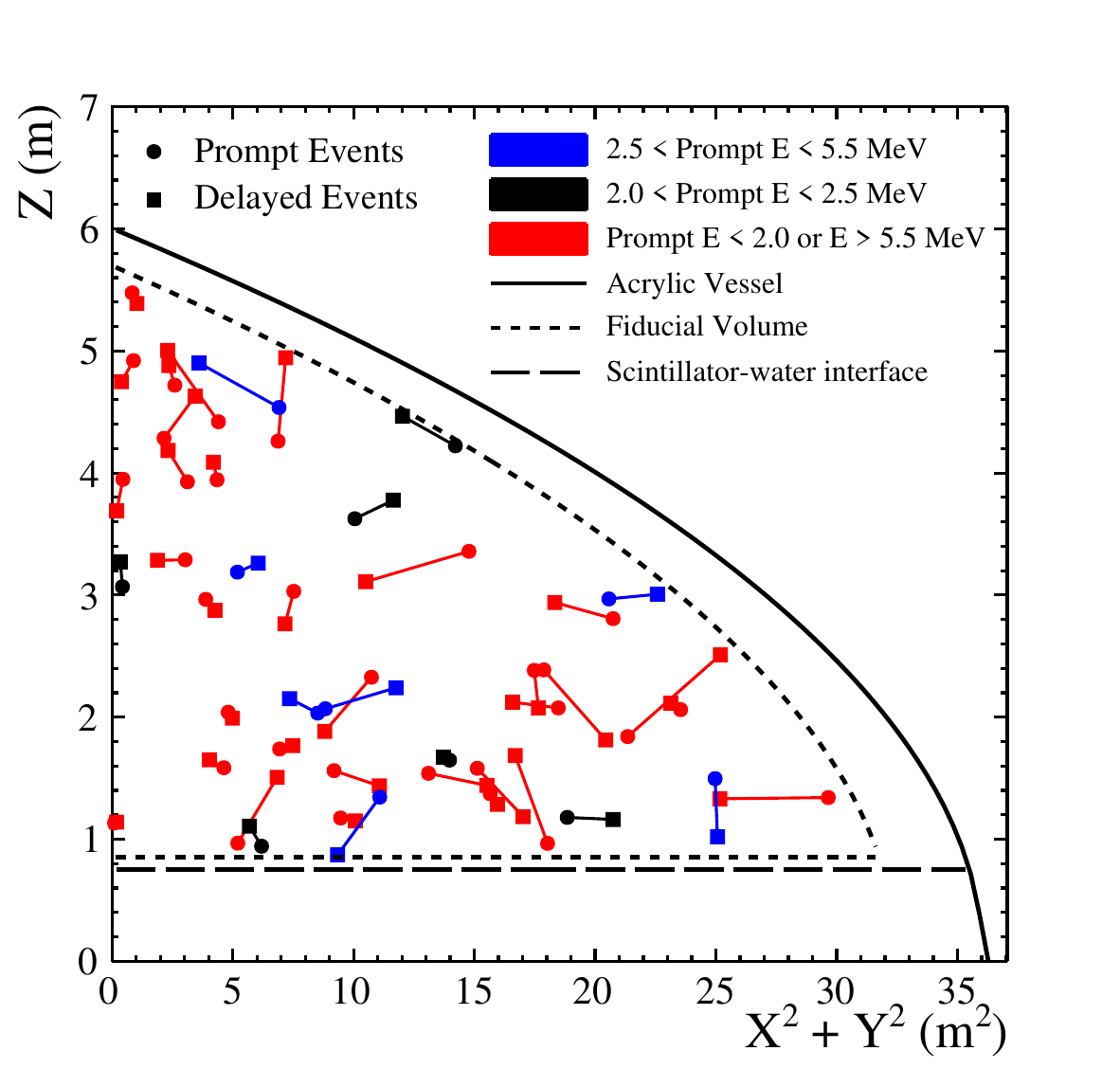}
\end{figure}

The expected numbers of signals and backgrounds are summarized in Table~\ref{tab:sum_sig_bkg}. 
The positions of the selected event pairs inside the detector are shown in Fig. \ref{fig:ZvsRho}. The event pairs are labeled in color according to the reconstructed energy of the prompt event: below 2 MeV and above 5.5 MeV (red), where mostly \alphan are expected; between 2.0 and 2.5 MeV (black); and between 2.5 and 5.5 MeV (blue), where reactor antineutrinos are expected to be dominant, as shown in Fig.~\ref{fig:Eprompt}.

\begin{table}[hbtp]\centering
\caption{Expected and fitted signal and background counts.\label{tab:sum_sig_bkg}} 
\begin{tabular}{lcc}
\hline
\hline
   &  Prediction & Fit result \\
\hline
Reactor IBD  & 9.5$\pm$0.3 & 9.5$\pm$0.3\\
Geo IBD & 2.2$\pm$2.2 & 2.5$\pm$2.1 \\
\alphan & 33.3$\pm$12.7 & 32.4$\pm$5.6\\
Sum         &  45.0$\pm$12.9  &  44.4$\pm$6.0 \\ \hline 
Observed    & \multicolumn{2}{c}{45}  \\
\hline
\hline
\end{tabular}
\end{table}

\subsection{Spectral analysis}

Figure~\ref{fig:Eprompt} shows the energy distribution of the selected prompt events. 
The impact of neutrino oscillation is most clearly observed between 2.5 MeV and 5.5 MeV, where the flux is most reduced by oscillations and the background counts are least.  

The prompt energy spectrum is fit with an extended $\log$ likelihood function to identify allowed regions of $\theta_{12}$ and $\Delta m^2_{21}$.  
In the fit, geoneutrinos are assigned a single spectrum, constructed with a U/Th ratio of 3.6 and with the average oscillation effect.  
Reactors at more than 1000~km are also represented by a single spectrum with average oscillation. 
The flux normalizations and energy-related systematic uncertainties are constrained in the fit as summarized in Table~\ref{tab:syst}. 

The systematic uncertainty on the number of target protons within the fiducial volume arises from 
differences in scintillator density between simulation and data, namely due to temperature fluctuations over time, and the uncertainty on the molecular composition, which sum to less than 1\%. 
Uncertainty in the position reconstruction translates to an uncertainty in the fiducial volume selection, which is estimated to be less than 1\%.  
A nonlinear uncertainty in the energy scale arising from Birks' law was tested and found to have a negligible impact on the fit results.  
All systematic uncertainties are much smaller than the statistical uncertainties from the data in this analysis.

\begin{table}[tbp]\centering
\caption{Systematic uncertainties and their 1$\sigma$ constraints in the fit.\label{tab:syst}}
\begin{tabular}{lc}
\hline \hline
Source & Constraint \\
\hline
Individual reactor rate  & 3\% \\
Geoneutrino rate & 100\% \\ 
\alphan ground state rate & 30\% \\
\alphan excited state rate & 100\% \\
\hline
Energy resolution & 3\% \\
Energy scale for $\beta$'s & 3\% \\
Energy scale for protons & 3\% \\
\hline \hline
\end{tabular}
\end{table}

At the best-fit point, the energy scale for the proton-scattering \alphan component is fit to be 2\% larger than the prediction, with a 3\% decrease in \alphan counts, and a 14\% increase in the geoneutrino flux, as shown in Table~\ref{tab:sum_sig_bkg}. These variations are all well within the constraints. 

\begin{figure}[htpb]
\caption{\label{fig:oscResult} Fitted likelihood value as a function of $\Delta m^2_{21}$ with $\theta_{12}$ fixed to the global result in Ref.~\cite{PDG2021}.} 
\includegraphics[width=\columnwidth]{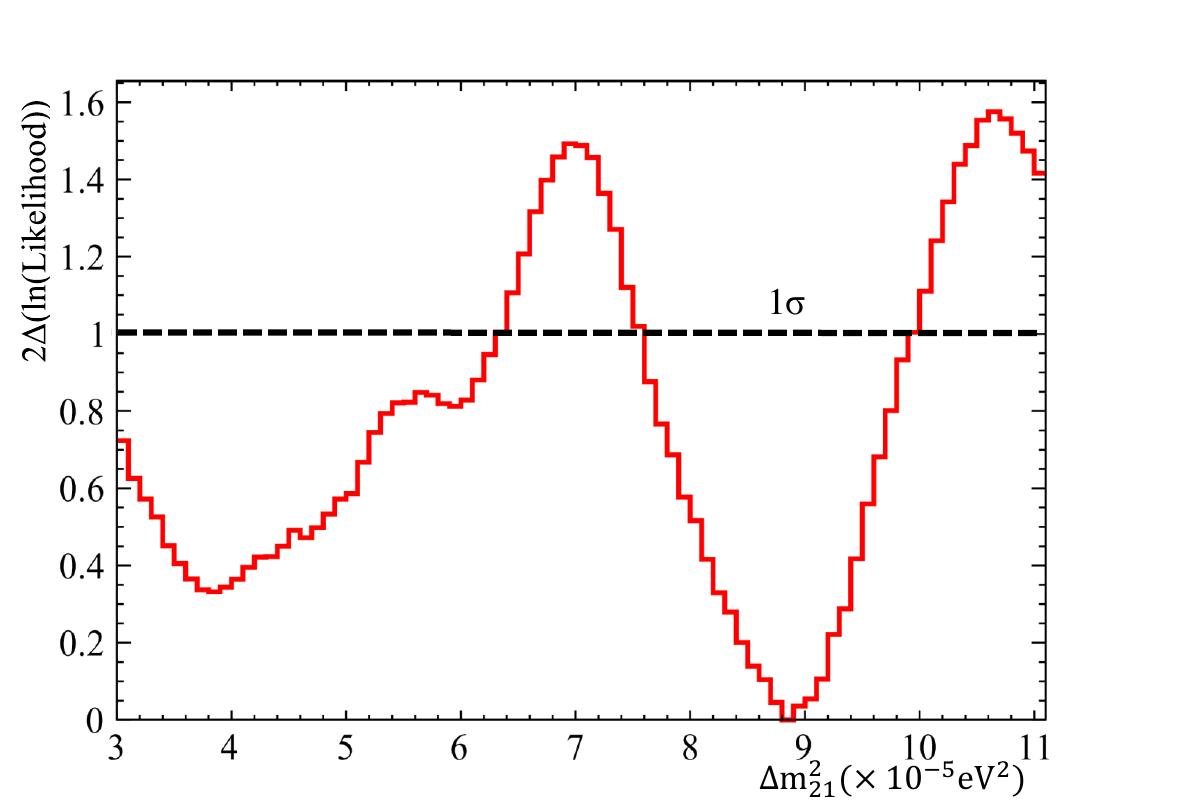}
\end{figure}

Since $\theta_{12}$ relates directly to counts while $\Delta m^2_{21}$ is very sensitive to the spectral shape, as illustrated by the different curves in Fig.~\ref{fig:EpromptPredict}, the large statistical uncertainty of the present dataset prevents a direct measurement of the mixing angle $\theta_{12}$ and so, it is fixed to the global average in Ref.~\cite{PDG2021} while fitting for $\Delta m^2_{21}$.  In the range allowed by previous measurements~\cite{PDG2021}, allowed regions for $\Delta m^2_{21}$ are identified at a 68\% confidence level, as shown in Fig.~\ref{fig:oscResult}. 
The best-fit is $\Delta m^2_{21}$ = (8.85$^{+1.10}_{-1.33}$) $\times$ 10$^{-5}$ eV$^2$. 
This result is favored compared to the likelihood fit without neutrino oscillation by a frequentist confidence level of 93.6\%.  It is consistent with the results from KamLAND and solar measurements at around 1$\sigma$.

The current result is combined with the global result, (7.53 $\pm$ 0.18) $\times$ 10$^{-5}$ eV$^2$ in Ref.~\cite{PDG2021}, by summing the 2$\Delta$($\ln$(Likelihood)) distribution in Fig.~\ref{fig:oscResult} with an assumed quadratic distribution whose width is set equal to 0.18 $\times$ 10$^{-5}$ eV$^2$ at 2$\Delta$($\log$(Likelihood)) = 1.  The resulting minimum value occurs at a $\Delta m^2_{21}$ that is 0.23\% larger than the global result, with an uncertainty that is 1\% smaller, giving (7.55 $\pm$ 0.18) $\times$ 10$^{-5}$ eV$^2$.  

\section{Future sensitivities}

The SNO+ detector has been acquiring data fully filled with 780 tonnes of scintillator since April 2022, and will continue to do so during the neutrinoless double beta decay search period, when the scintillator will be loaded with tellurium. Below, we describe the prospects for future sensitivities.

\subsection{\alphan background}\label{subsec:an}

The dominant ground state signal of \alphan events has prompt energies below 3.5~MeV (see Fig.~\ref{fig:Eprompt}) and produces scintillation photons primarily from multiple protons scattered by the neutron on the scale of a nanosecond.  This additional smearing in the photon time profile provides a strong handle for the pulse shape discrimination of $n$'s and $\beta$'s.  
Thus, a likelihood ratio is calculated by comparing the corrected hit times of prompt events below 3.5~MeV to PDFs of those from simulated \alphan and IBDs.  More details can be found in Ref.~\cite{CM2022}.  A cut on this ratio at 0.0 reduces the \alphan expectation by 70\% while keeping 93\% of reactor IBDs. In the data, 20 out of the 45 observed coincidences survive and fit to 8.9 reactor IBDs, 2.2 geo IBDs, and 7.2 \alphan events. 

The sensitivity of the oscillation fit is not improved with this purer preliminary sample due to the very limited signal statistics of the present dataset.  
An improved implementation of this event discriminator will be used to reduce the impact of the \alphan background in future measurements.  

After the detector was fully filled with scintillator, measurements of the specific activity of $^{210}$Po showed a rate about five times lower than the present dataset, implying a factor of five reduction in the \alphan background.  
One hypothesis for this decrease is that $^{210}$Po from the inner surfaces of the scintillator plant piping (and perhaps the neck of the acrylic vessel) was initially released into the flow as scintillator filling began.  Subsequently, the amount of contamination introduced per volume of scintillator added would decrease as filling progressed.  
The observed time evolution of the $^{210}$Po background saw it decaying slightly slower than its 138-day halflife.  Levels of $^{210}$Bi (supported by $^{210}$Pb) were seen to be relatively constant in comparison, with a specific activity that was 10-20 times lower than its daughter $^{210}$Po.  Therefore, this out-of-equilibrium $^{210}$Po initial injection model is consistent with observations.  
With scintillator filling completed in 2022, the lower specific activity of $^{210}$Po is expected to remain constant.  

Upon loading tellurium into the scintillator, the light yield is expected to decrease~\cite{Auty:2022lgh}, but not degrade the energy resolution enough to impact this analysis.  After the present dataset, the concentration of the fluor PPO was increased to 2.2~g/L, and later, 2.2~mg/L of the wavelength-shifter bis-MSB was added to the scintillator, both of which have significantly increased the light collected.  
With the addition of a few tonnes of $^\mathrm{nat}$Te and related chemicals, some increase in radioactive backgrounds can be expected, possibly resulting in an increase in the main \alphan background, as well as in random coincidences.  As mitigation, we have developed a powerful event discriminator, a purification technique and the infrastructure to implement it, and the telluric acid to be used has been stored 2 km underground since 2015, allowing a number of cosmogenically-induced isotopes to reduce in population by decay.

\subsection{Neutrino oscillation parameters}

\begin{figure}[tbp]
\caption{\label{fig:oscFuture} Predicted uncertainty on $\Delta m^2_{21}$ vs. livetime.  The dashed line shows the uncertainty from the current best measurement from KamLAND~\cite{KamLAND:2013rgu}.  The solid blue curve assumes the \alphan rate measured in the full fill detector and the solid red curve assumes that there is no \alphan background.}
\includegraphics[width=\columnwidth]{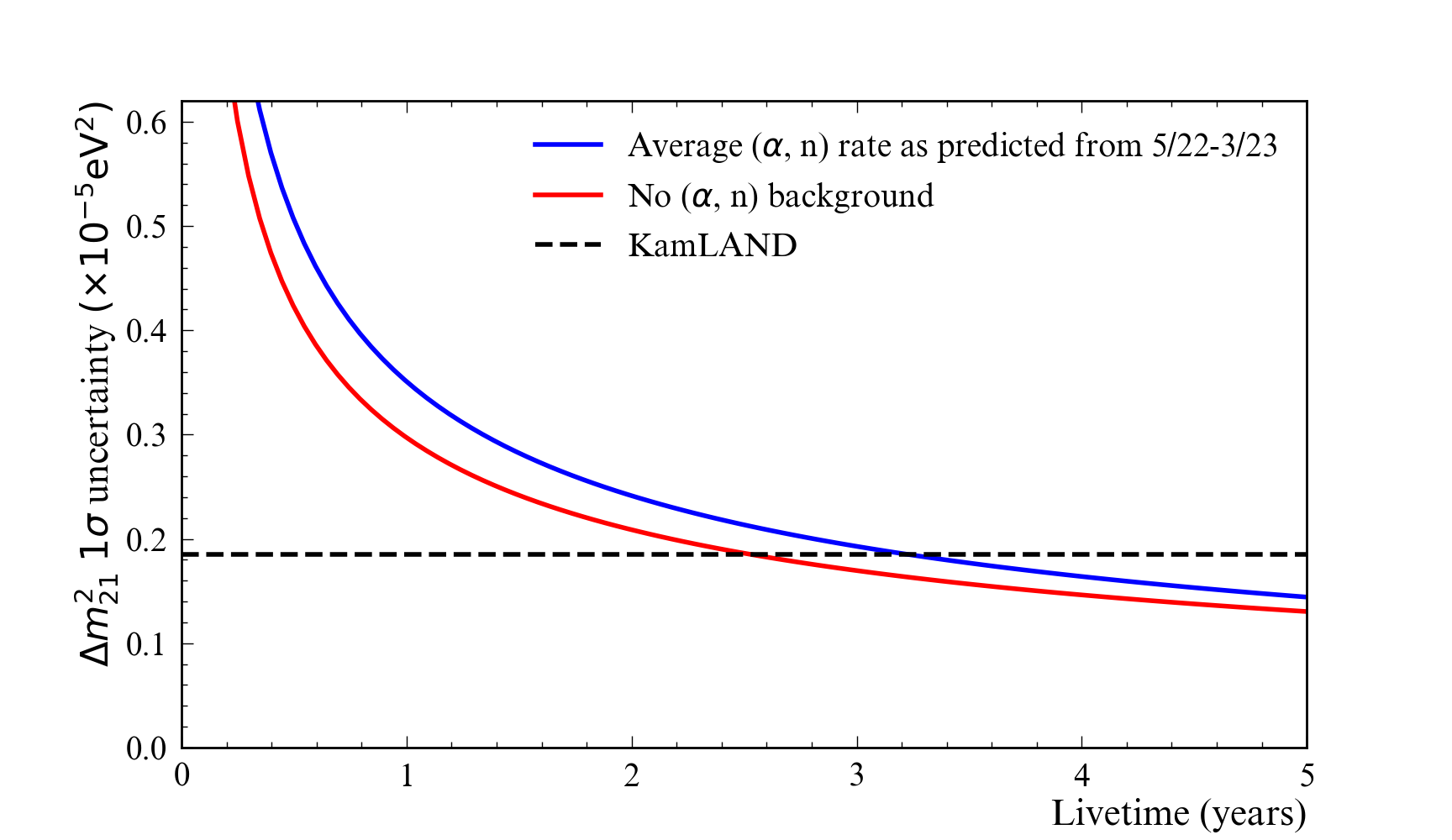}
\end{figure}

Figure~\ref{fig:oscFuture} shows the expected evolution of the uncertainty on $\Delta m^2_{21}$ measured by the SNO+ experiment as a function of livetime.  
The projection assumes a fully-filled detector and the observed specific activity of $^{210}$Po that is five times lower than that observed for the present analysis.  With the resulting \alphan rate, SNO+ expects to surpass the present best measurement after collecting about 3.3~years of data. 
The event discriminator described above will reduce the impact of the \alphan background, which could ideally allow this result to be achieved in as little as 2.6~years.  
A reduced \alphan background would also increase the sensitivity to $\theta_{12}$.

\subsection{Geoneutrino flux}

The current dataset does not allow a measurement of the geoneutrino flux due to the limited statistics and overlapping \alphan energy spectrum. 
The SNO+ experiment will measure the total geoneutrino flux with enhanced analytical methods and a reduced \alphan background, at an expected rate of about 20 selected geo IBDs per year of livetime in the fully-filled detector. 
Such a measurement is expected to contribute significantly to the global analysis of Earth models in conjunction with previous geoneutrino measurements at other locations~\cite{geoKL,geoBX}.

\section{Conclusion}
With 125.4 days of data and 320 tonnes of scintillator with approximately 0.6~g/L of PPO, the SNO+ collaboration has measured the oscillation of antineutrinos from distant nuclear reactors. 
The statistical uncertainty prevents a measurement of the mixing angle $\theta_{12}$, which is therefore fixed to the global result when fitting for $\Delta m^2_{21}$.  In the ranges of values allowed by previous measurements of solar neutrinos and the lone measurement of long-baseline reactor antineutrinos from KamLAND, the resulting likelihood curve is compatible with the previous measurements and produces a best fit of $\Delta m^2_{21}$ = (8.85$^{+1.10}_{-1.33}$) $\times$ 10$^{-5}$ eV$^2$. 

The detector has since completed filling and now holds 780 tonnes of liquid scintillator.  
The loading of PPO was completed in April 2022, reaching a concentration of 2.2~g/L.  
Measurements of the $^{210}$Po specific activity, which determines the \alphan background rate, show a decrease by a factor of roughly five compared with the partial-fill data used in the current analysis.  

The precision of $\Delta m^2_{21}$ from the SNO+ experiment alone is expected to surpass that from the present global result after about three years of data acquisition with the full detector.  
SNO+ will also provide a measurement of oscillation angle $\theta_{12}$ and a first measurement of the U/Th geoneutrino flux in the North American Plate and in the Western Hemisphere.

\begin{acknowledgments}
Capital funds for SNO\raisebox{0.5ex}{\tiny\textbf{+}} were provided by the Canada Foundation for Innovation and matching partners: 
Ontario Ministry of Research, Innovation and Science, 
Alberta Science and Research Investments Program, 
Queen’s University at Kingston, and 
the Federal Economic Development Agency for Northern Ontario. 
This research was supported by 
{\it Canada: }
the Natural Sciences and Engineering Research Council of Canada, 
the Canadian Institute for Advanced Research, 
the Ontario Early Researcher Awards, 
the Arthur B. McDonald Canadian Astroparticle Physics Research Institute; 
{\it U.S.: }
the Department of Energy (DOE) Office of Nuclear Physics, 
the National Science Foundation and the DOE National Nuclear Security
Administration through the Nuclear Science and Security Consortium; 
{\it UK: }
the Science and Technology Facilities Council and the Royal Society; 
{\it Portugal: } 
Funda\c{c}\~{a}o para a Ci\^{e}ncia e a Tecnologia (FCT-Portugal); 
{\it Germany: }
the Deutsche Forschungsgemeinschaft; 
{\it Mexico: }
DGAPA-UNAM and Consejo Nacional de Ciencia y Tecnolog\'{i}a; 
{\it China: }
the Discipline Construction Fund of Shandong University.  
We also thank SNOLAB and SNO\raisebox{0.5ex}{\tiny\textbf{+}} technical
staff; the Digital Research Alliance of Canada; the
GridPP Collaboration and support from Rutherford
Appleton Laboratory; and the Savio computational cluster
at the University of California, Berkeley. Additional long-term
storage was provided by the Fermilab Scientific Computing
Division.

For the purposes of open access, the authors have applied a Creative Commons Attribution licence to any Author Accepted Manuscript version arising. Representations of the data relevant to the conclusions drawn here are provided within this paper.
\end{acknowledgments}

\appendix*
\section{Event information}\label{sec:eventInfo}
Table~\ref{tab:event_info} provides information about every event selected by the IBD selection criteria described in Sec.~\ref{sec:sigBg}.  

\begin{table*}[ht!]
    \centering
    \caption{Information about the 45 selected coincidence pairs.}  
    \label{tab:event_info}
    \begin{tabular}{ccc|ccc|ccc} \hline \hline 
         \multicolumn{3}{c|}{Prompt} & \multicolumn{3}{c|}{Delayed} & \\ \hline 
         Energy [MeV] & $R$ [m] & $Z$ [m] & Energy [MeV] & $R$ [m] & $Z$ [m] & Time diff. [$\mu$s] & Position diff. [mm] & Date \\ \hline 
2.36 & 5.66 & 4.22 & 1.89 & 5.65 & 4.47 & 143.2 & 394 & 2020/03/30\\
4.75 & 3.92 & 3.19 & 2.19 & 4.09 & 3.26 & 116.0 & 201 & 2020/04/01\\
3.13 & 5.24 & 4.54 & 1.91 & 5.26 & 4.90 & ~39.4 & 818 & 2020/04/01\\
2.44 & 4.09 & 1.65 & 1.96 & 4.06 & 1.67 & 127.6 & ~42 & 2020/04/01\\
1.76 & 3.93 & 1.54 & 1.91 & 4.19 & 1.44 & 251.5 & 402 & 2020/04/01\\
1.66 & 3.16 & 1.74 & 1.89 & 3.26 & 1.77 & ~~8.3 & 121 & 2020/04/04\\
1.79 & 4.02 & 2.33 & 2.00 & 3.52 & 1.88 & 309.5 & 636 & 2020/04/08\\
1.67 & 5.55 & 5.48 & 1.89 & 5.48 & 5.39 & ~90.6 & 260 & 2020/04/11\\
1.33 & 3.29 & 1.17 & 2.06 & 3.38 & 1.15 & 796.8 & 428 & 2020/04/18\\
1.13 & 4.86 & 2.39 & 2.12 & 4.87 & 1.81 & ~17.1 & 719 & 2020/04/18\\
1.37 & 2.67 & 1.59 & 2.01 & 2.60 & 1.65 & ~21.5 & 189 & 2020/04/22\\
1.11 & 5.00 & 5.62 & 2.28 & 4.26 & 4.94 & 678.4 & 688 & 2020/04/25\\
2.22 & 4.50 & 1.18 & 2.02 & 4.70 & 1.16 & 445.9 & 290 & 2020/04/30\\
1.97 & 4.99 & 4.72 & 2.20 & 5.12 & 4.88 & 160.6 & 303 & 2020/06/24\\
1.10 & 4.97 & 1.84 & 2.27 & 5.61 & 2.51 & ~12.1 & 964 & 2020/06/26\\
1.20 & 4.19 & 1.37 & 2.05 & 4.20 & 1.28 & ~78.6 & ~99 & 2020/06/29\\
1.25 & 5.27 & 2.06 & 2.15 & 5.25 & 2.11 & ~40.1 & 166 & 2020/07/08\\
0.99 & 4.36 & 0.97 & 2.17 & 4.42 & 1.69 & 507.7 & 854 & 2020/07/10\\
2.23 & 2.66 & 0.94 & 2.19 & 2.63 & 1.11 & 296.9 & 356 & 2020/07/16\\
1.17 & 4.46 & 3.94 & 2.09 & 4.57 & 4.09 & 241.1 & 302 & 2020/07/24\\
0.91 & 5.01 & 4.92 & 2.24 & 4.79 & 4.75 & 138.0 & 521 & 2020/07/24\\
1.14 & 3.41 & 1.56 & 2.09 & 3.62 & 1.44 & 276.1 & 486 & 2020/08/06\\
1.15 & 4.20 & 1.58 & 2.14 & 4.29 & 1.19 & ~49.0 & 531 & 2020/08/09\\
1.60 & 3.72 & 3.29 & 2.07 & 3.56 & 3.28 & 218.5 & 368 & 2020/08/15\\
2.80 & 5.22 & 1.50 & 1.86 & 5.11 & 1.02 & ~99.6 & 498 & 2020/08/19\\
0.97 & 5.61 & 1.34 & 2.34 & 5.19 & 1.33 & 586.7 & 445 & 2020/08/21\\
1.45 & 4.09 & 3.03 & 2.23 & 3.85 & 2.76 & ~47.1 & 587 & 2020/08/23\\
5.74 & 1.18 & 1.13 & 2.02 & 1.22 & 1.14 & 182.0 & 123 & 2020/08/26\\
1.25 & 2.48 & 0.97 & 2.03 & 3.02 & 1.51 & 518.9 & 648 & 2020/08/26\\
1.11 & 4.31 & 3.93 & 2.25 & 4.45 & 4.18 & ~~8.2 & 358 & 2020/09/01\\
2.33 & 4.82 & 3.63 & 2.06 & 5.09 & 3.78 & 218.3 & 376 & 2020/09/01\\
2.00 & 3.14 & 3.07 & 2.21 & 3.33 & 3.27 & ~66.8 & 314 & 2020/09/04\\
4.05 & 3.62 & 2.07 & 2.08 & 4.10 & 2.24 & 345.8 & 497 & 2020/09/05\\
1.43 & 5.35 & 2.81 & 2.20 & 5.19 & 2.94 & 372.8 & 310 & 2020/09/12\\
1.45 & 4.81 & 2.38 & 2.08 & 4.69 & 2.08 & ~90.8 & 401 & 2020/09/12\\
1.81 & 4.01 & 3.95 & 2.07 & 3.72 & 3.69 & 352.7 & 532 & 2020/09/19\\
1.21 & 5.10 & 3.36 & 2.27 & 4.49 & 3.11 & 694.3 & 705 & 2020/09/25\\
1.14 & 3.56 & 2.96 & 2.10 & 3.54 & 2.88 & 274.2 & 371 & 2020/10/02\\
1.06 & 4.89 & 4.42 & 2.11 & 5.23 & 5.01 & ~~9.4 & 858 & 2020/10/04\\
2.82 & 5.42 & 2.97 & 2.20 & 5.62 & 3.01 & ~49.3 & 309 & 2020/10/08\\
1.41 & 4.77 & 2.08 & 2.05 & 4.59 & 2.12 & ~~4.2 & 234 & 2020/10/11\\
1.08 & 4.53 & 4.28 & 2.18 & 4.99 & 4.63 & 152.0 & 552 & 2020/10/19\\
1.53 & 3.00 & 2.04 & 2.17 & 2.99 & 1.99 & 411.6 & 127 & 2020/10/19\\
4.93 & 3.56 & 2.03 & 2.14 & 3.46 & 2.15 & ~31.4 & 296 & 2020/10/21\\
3.39 & 3.59 & 1.34 & 2.20 & 3.18 & 0.87 & 225.5 & 647 & 2020/10/21\\
\hline \hline

    \end{tabular}
\end{table*}

\clearpage


\end{document}